\documentstyle[11pt,epsf,multicol,amsmath,amssymb]{article}

\begin{document}

\newlength{\smallestlength}
\setlength{\smallestlength}{-0.05in}

\newlength{\smalllength}
\setlength{\smalllength}{-0.1in}

\newlength{\midlength}
\setlength{\midlength}{-0.15in}

\newlength{\middlelength}
\setlength{\middlelength}{-0.2in}

\newlength{\biglength}
\setlength{\biglength}{-0.25in}

\newlength{\largelength}
\setlength{\largelength}{-0.38in}

\setcounter{page}{1}
\pagenumbering{arabic}

\title{A comparison of jamming behavior in systems composed of dimer-  
and ellipse-shaped particles}

\author{
  Carl F. Schreck$^{1}$, Ning Xu$^{2}$, and Corey S. O'Hern$^{3,1}$\\    
$^{1}$Department
of Physics, Yale University, New Haven CT 06520-8120\\
$^{2}$Department of Physics, University of Science and Technology,\\ 
Hefei 230026, China\\
$^{3}$Department of Mechanical Engineering, Yale University, \\ New
Haven, CT 06520-8286
} 

\maketitle

\begin{abstract}
We compare the structural and mechanical properties of static packings
composed of frictionless convex (ellipses) and concave (rigid dimers)
particles in two dimensions. We employ numerical simulations to
generate static packings and measure the shear stress in response to
applied simple shear strain as a function of the aspect ratio and
amount of compression.  We find that the behavior near jamming is
significantly different for ellipses and dimers even though both
shapes are roughly characterized by the aspect ratio and possess the
same number of translational and rotational degrees of freedom per
particle.  For example, we find that ellipse packings are hypostatic
(not isostatic as found for dimers), display novel power-law scaling
of the static linear shear modulus and contact number with the amount
of compression, and possess stress-strain relations that are 
qualitatively different from that for dimers.  Thus,
we observe that important macroscopic properties of static packings of
anisotropic particles can depend on the microscale geometrical
features of individual particles.
\end{abstract}

\section{Introduction}
\label{intro}

Significant progress has been made in understanding the jamming
transition that occurs in collections of frictionless spherical
particles with purely repulsive short-range interactions, where
systems transform from liquid- to solid-like states as a function of
increasing packing fraction or decreasing applied shear
stress~\cite{lucid,newlucid,van}.  Key findings include the power-law
scaling of the static shear modulus with packing fraction above the
onset of jamming~\cite{Ohern2003,Ellenbroek2006}, the identification
of a growing lengthscale as the system approaches the jamming
transition~\cite{silbert}, above which the system can be described as
an elastic material~\cite{Ellenbroek2009}, and an abundance of
low-energy excitations in the density of vibrational
modes~\cite{wyart}.  Much of this behavior stems from the fact that
frictionless, static packings of spherical particles are typically
isostatic since they possess the minimal number of contacts per
particle $z_{\rm iso} = 2d$, where $d$ is the spatial dimension,
required for mechanical stability~\cite{Witten2000}.

\begin{figure}
\epsfxsize=5.0in
\centerline{\epsffile{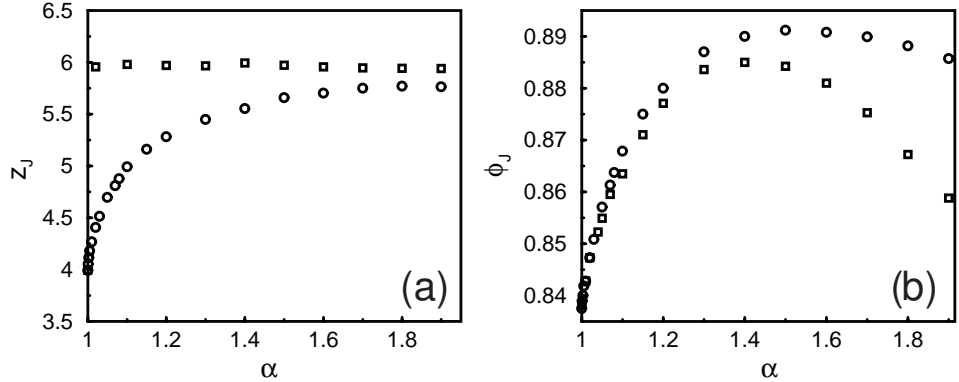}}
\caption{Ensemble averaged (a) contact number $z_J$ and (b) packing fraction
$\phi_J$ at jamming as a function of aspect ratio $\alpha$ for dimers
(squares) and ellipses (circles) for $N=480$ particles.
\label{fig:zphi_dimer_ellipseCS2}}
\vspace{-0.0in}
\end{figure}

However, less progress has been made in understanding the jamming
transition in particulate systems composed of nonspherical particles,
despite the fact that these systems display striking
mechanical~\cite{franklin} and rheological~\cite{mori} properties, and
are more relevant for industrial applications and in nature.  An
important difference between static packings of frictionless spherical
versus ellipsoidal particles is that the latter are typically
hypostatic, not isostatic, with fewer contacts than required to
constrain all of the translational and rotational degrees of freedom
using straightforward counting
arguments~\cite{Donev2007,Zeravcic2009}.  Previous studies have found
that for ellipse packings the contact number at jamming $z_J < z_{\rm
iso} = 2 d_f$, where $d_f=3$ in two dimensions ($2D$), over a wide
range of aspect ratios $\alpha$~\cite{Donev2007}.  We include similar
results from our simulations of static ellipse packings in
Fig.~\ref{fig:zphi_dimer_ellipseCS2} (a). $z_J < z_{\rm iso}$ for
small aspect ratios, but slowly approaches a value $z_J^*$ that is only a few
percent below the isostatic value as $\alpha$ increases.  The packing
fraction at jamming $\phi_J$ for ellipses, shown in panel (b),
possesses a peak, which is only a few percent lower than the
crystalline value for spherical particles, near $\alpha \sim
1.5$~\cite{Donev2007,chaikin}.

\section{Motivation}
\label{motive}

In this manuscript, we investigate the generality of these results for
the behavior near jamming of frictionless, anisotropic particles by
comparing the structural and mechanical properties of two classes of
nonspherical shapes: {\it convex} (ellipses) and {\it concave} (rigid
dimers) particles.  We find that the behavior near jamming for rigid
dimers differs significantly from that for ellipses even though both
shapes are roughly characterized by the aspect ratio and
possess the same number of translational and rotational degrees of
freedom per particle.

We find several key differences between the structural and mechanical
properties of static packings of dimers and ellipses.  First, our
simulations indicate that static packings of dimers are {\it
isostatic} (not hypostatic as found for ellipses) with $z\simeq z_{\rm
iso}$ contacts per particle over the full range of aspect ratios
studied as shown in Fig.~\ref{fig:zphi_dimer_ellipseCS2}
(a)~\footnote{It is still an open question which concave particle shapes yield
isostatic packings, and which do not.}. Second, ellipse packings display novel
power-law scaling of the static linear shear modulus $G$ and contact
number $z-z_J$ with $\phi-\phi_J$~\cite{Mailman2009}---both scale
linearly with $\phi-\phi_J$.  In contrast, for dimer packings $G$ and
$z-z_J$ scale as $(\phi-\phi_J)^{0.5}$, which is the same scaling
found for static packings of spherical particles~\cite{Ohern2003}.
Third, we find that the shear stress-strain relations for packings of
dimers and ellipses are qualitatively different.  For example, at
large compressions, the stress response (below the yield stress) to
applied strain depends strongly on aspect ratio for ellipses, but it
is nearly independent of aspect ratio for dimers. Also, at small
compressions, dimer packings display nearly perfect plastic response
in a region of strain where ellipse packings possess roughly linear
response.

This manuscript is organized as follows.  In
Sec.~\ref{methods} we describe the computational methods for
generating static packings of ellipses and dimers and then applying
quasistatic simple shear to measure the mechanical response.  In
Sec.~\ref{results}, we present our results for the linear static shear
modulus, contact number, stress-strain relation, and particle rearrangement
statistics. In Sec.~\ref{future}, we discuss our conclusions and
identify possible future research directions.  We also include
five appendices, which provide the details necessary for calculating the
packing fraction for dimers, contact distance between ellipses, and
forces, torques, and stress tensor for anisotropic
particles~\cite{edwards}.

\begin{figure}
\epsfxsize=4.0in
\centerline{\epsffile{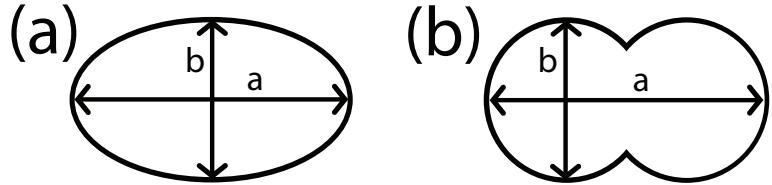}}
\vspace{-0.0in}
\caption{Definition of the aspect ratio $\alpha = a/b$ (ratio of the
major to minor axes) for (a) ellipses and (b) dimers.
\label{fig:ellipsedimer}}
\vspace{-0.0in}
\end{figure}

\section{Computational methods}
\label{methods}

We performed computational studies to measure the structural and
mechanical properties of static packings of rigid dimers and
ellipse-shaped particles in $2D$. The particle shapes are shown in
Fig.~\ref{fig:ellipsedimer}. The rigid dimers are formed by fusing
identical disks together.  We study aspect ratios $\alpha = a/b$ in
the range $1 \le \alpha \le 2$, where $a$ and $b$ are the length of
the major and minor axes, respectively.  To inhibit crystallization,
we focus on bidisperse mixtures of particles: $2 N/3$ particles
with minor axis $b$ and $N/3$ larger particles with minor axis $1.4
b$.  The particles are enclosed in square simulation cells with box
length $L$ and periodic boundary conditions.  System sizes were varied from
$24 \le N \le 480$.

The particles interact via soft, pairwise, purely repulsive linear spring
potentials.  The total potential energy is therefore given by 
\begin{equation}
\label{interaction}
V = \sum_{i>j} V\left(\frac{r_{ij}}{\sigma_{ij}}\right) = \frac{\epsilon}{2} \sum_{i>j} \left( 1 - \frac{r_{ij}}{\sigma_{ij}}
\right)^{2} \Theta \left( 1 - \frac{r_{ij}}{\sigma_{ij}} \right), 
\end{equation}
where ${\vec r}_{ij}$ is the vector separation between the centers of
particles $i$ and $j$, $\epsilon$ is the characteristic energy scale
of the interaction, $\Theta(x)$ is the Heaviside function, and
$\sigma_{ij}$ is the contact distance that in general depends on the
orientation of particles $i$ and $j$, ${\hat \mu}_i$ and ${\hat
\mu}_j$, and ${\hat r}_{ij}$.

\paragraph{Contact distance}
Determining the interactions between dimers is straightforward: one
can identify overlaps between individual disks (monomers) on different dimers.
Thus, the contact distance between disk $i$ on a given dimer and disk
$j$ on a different dimer is $\sigma_{ij} = (b_i + b_j)/2$, and the
total potential energy can be obtained by summing up the contributions
$V(r_{ij}/\sigma_{ij})$ over all disk-disk interactions for disks on
distinct dimers.

\begin{figure}
\epsfxsize=4.5in
\centerline{\epsffile{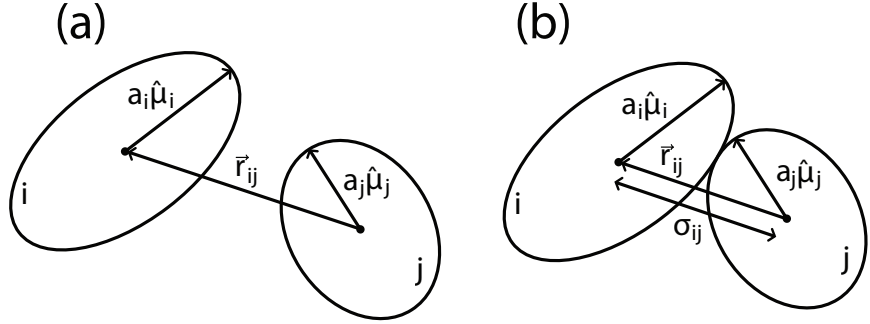}}
\vspace{-0.0in}
\caption{(a) Ellipses $i$ and $j$ with orientations ${\hat \mu}_i$ and
${\hat \mu}_j$ at center-to-center separation ${\vec r}_{ij}$. (b) The
contact distance $\sigma_{ij}$ is obtained by identifying the 
point of contact when the two ellipses are brought together at fixed
orientation.
\label{fig:contactdistance}}
\vspace{-0.0in}
\end{figure}

The contact distance $\sigma_{ij}$ between ellipses is more difficult
to calculate than that for dimers.  We define $\sigma_{ij}$ as the
distance at which two ellipses will first come into contact when moved
along their center-to-center direction while their orientations are
held fixed.  Fig.~\ref{fig:contactdistance} illustrates how
$\sigma_{ij}$ is measured for ellipses $i$ and $j$ with orientations
${\hat \mu}_i$ and ${\hat \mu}_j$ at separation ${\vec r}_{ij}$.  We
calculate the contact distances $\sigma_{ij}$ analytically in systems
of bidisperse ellipses using the Perram-Wertheim
formulation~\cite{Cleaver,Perram,Perram1996,Schreck2009}. Further
details are provided in Appendix~\ref{contactdistance}.
 
\begin{figure}
\epsfxsize=4.0in
\centerline{\epsffile{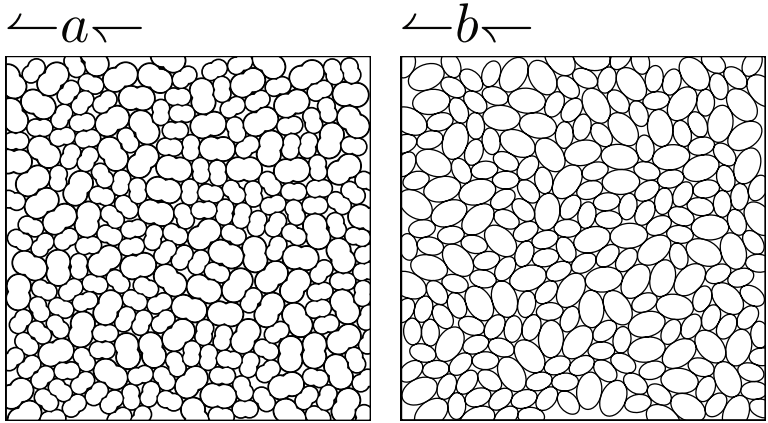}}
\vspace{-0.0in}
\caption{Snapshots of static packings of $N=240$ bidisperse (a)
rigid dimers and (b) ellipses with $\alpha=1.5$.
\label{fig:n240_2D_systems}}
\vspace{-0.0in}
\end{figure}

\paragraph{Packing-generation algorithm}
We generate static, zero-pressure packings of bidisperse dimers and
ellipses using a generalization of the compression/decompression
method employed in our previous studies of spherical
particles~\cite{gao1,gao2}.  Representative packings of dimers and
ellipses are shown in Fig.~\ref{fig:n240_2D_systems}.  We briefly
outline the packing-generation procedure here for completeness.

We begin the packing-generation process by choosing random initial
particle positions and orientations within the simulation cell at
packing fraction $\phi_0 = 0.20$ (which is well below the minimum
packing fraction at which frictionless packings of ellipses and dimers
occur in $2D$ for $1 \le \alpha \le 2$).  We successively increase or
decrease the minor axes of the particles while maintaining the aspect
ratio, with each compression or decompression step followed by
conjugate gradient minimization~\cite{numrec} of the total energy
(\ref{interaction}). The system is decompressed when the total energy
at a local minimum is nonzero---{\it i.e.}, there are finite particle
overlaps.  If the potential energy of the system is zero and gaps
exist between particles, the system is compressed.  The increment by
which the packing fraction $\phi$ is changed at each compression or
decompression step is gradually decreased. The process is stopped when
the total potential energy per particle $V/\epsilon N \ll 1$.
Further details of the packing-generation algorithm are provided in
Appendix~\ref{packing}.

The packing fraction $\phi_J$, contact number $z_J$, and mechanical 
response are used to characterize each static packing.  The
packing-generation process is repeated at least $100$ times at each
$\alpha$ to generate configurational averages.  Once the packings at
jamming onset are generated, they can be successively compressed by
small amounts $\Delta \phi$, followed by energy minimization at each
step, to yield sets of configurations at fixed $\phi-\phi_J$.

\paragraph{Quasistatic simple shear}
To determine the mechanical properties of static packings of dimers
and ellipses, we studied their response to quasistatic simple shear at
fixed area.  We first initialized the system with an unstrained
packing at a given $\phi-\phi_J$ and successively applied to each
particle $i$ small affine simple shear strain steps $\delta
\gamma=\delta x/L$ along the $x$-direction with a gradient in the
$y$-direction:
\begin{eqnarray}
\label{strain}
x_i \rightarrow x_i + \delta \gamma y_i,
\end{eqnarray}
where ${\vec r}_i=(x_i,y_i)$ is the location of the center of mass of
particle $i$.  To be consistent with simple shear, at each strain
step, the angle $\theta_i = \cos^{-1} ({\hat \mu}_i \cdot {\hat x})$ that particle $i$
makes with the $x$-axis was also rotated:
\begin{equation}
\label{angle}
\theta_i \rightarrow \cot^{-1}(\cot \theta_i +\delta \gamma).
\end{equation}
Each shear strain step was followed by conjugate gradient energy
minimization using Lees-Edwards (shear periodic) boundary
conditions~\cite{allen}. For most studies, $\delta \gamma = 10^{-3}$
with accumulated strains $\gamma_t = 1$.  We verified that smaller
strain steps yielded similar results.  During the quasistatic shear,
we measured the shear stress (in units of $\epsilon/b$), contact
number, and statistics of particle rearrangement events.  These
measurements are described in the following Sec.~\ref{results}.
Details of the stress calculations are provided in
Appendices~\ref{torque} and~\ref{shear}.

\section{Results and discussion}
\label{results}

We present several measurements of the structural and mechanical
properties of ellipse and dimer packings as a function of aspect ratio
and compression $\phi-\phi_J$ including the contact number, shear
modulus, yield stress, and other features of the stress-strain
relations.

\begin{figure}
\epsfxsize=3.5in
\centerline{\epsffile{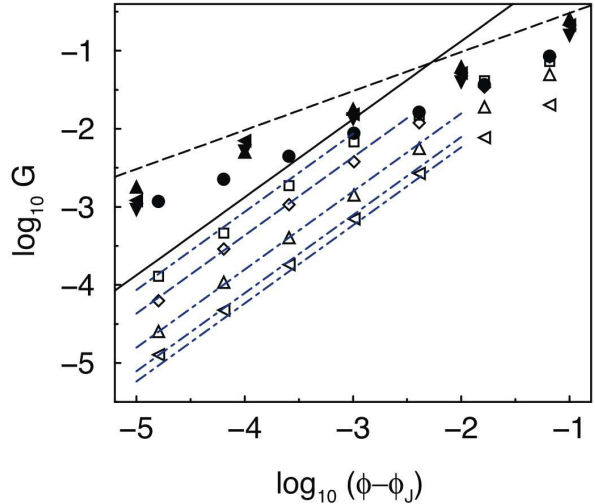}}
\vspace{-0.1in}
\caption{Static shear modulus $G$ versus $\phi-\phi_J$
for $N=480$ ellipses (open symbols) and $240$ dimers (filled symbols) at
$\alpha = 1.0$ (circles), $1.002$ (squares), $1.01$ (diamonds), $1.05$
(upward triangles), $1.1$ (leftward triangles), $1.5$ (downward
triangles), and $2.0$ (rightward triangles).  The solid (dashed) line
has slope $1$ ($0.5$).  The dot-dashed lines have the form $G= 0.6
(\phi-\phi_J)/(\alpha-1)^{0.44}$.
\label{fig:G_dimers_ellipses}}
\vspace{-0.0in}
\end{figure}

\paragraph{Contact number at jamming} The contact number is defined by 
$z = N_c/(N-N_r)$, where $N_c$ is the number of contacts
(interparticle overlaps) in the packing. $N_r$ is the number of
rattler particles with fewer than three contacts.  The contact network
is found by identifying all interparticle contacts, and then
recursively removing rattler particles until there are none remaining
in the packing.

In Fig.~\ref{fig:zphi_dimer_ellipseCS2} (a) we show results for the
contact number $z_J$ at jamming for ellipse and dimer packings.  We
find that ellipse packings are hypostatic with $z_J < 2 d_f$ over the
range of aspect ratio $1 \leq \alpha \leq 2$, while dimer packings are
isostatic with $z_J \simeq 2d_f$ over the same range of $\alpha$.  We
showed previously~\cite{Mailman2009,Schreck2009} that
hypostatic packings of ellipsoidal particles possess vibrational modes
that are quartically (not quadratically) stabilized with the number of
quartic modes determined by the deviation from isostaticity, $z_{\rm
iso} - z_J$.  In contrast, all vibrational modes for dimer packings are
quadratically stabilized since they are isostatic.

\begin{figure}
\epsfxsize=3.5in
\centerline{\epsffile{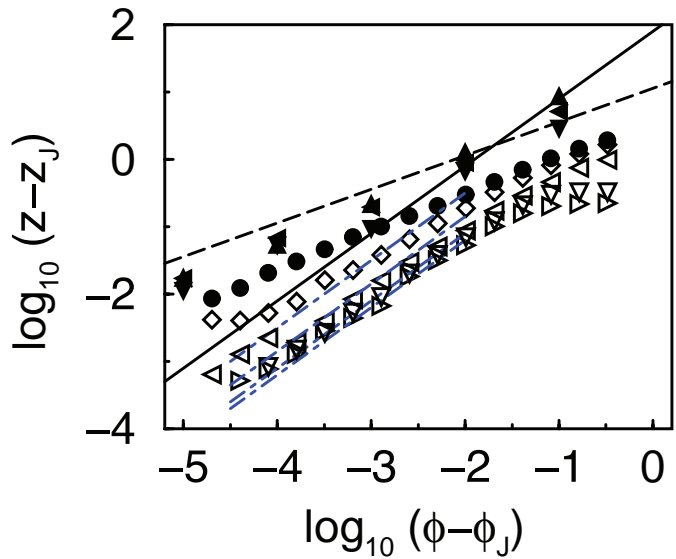}}
\vspace{-0.2in}
\caption{Deviation in the contact number $z$ from the value at 
jamming $z_J$ versus $\phi-\phi_J$
for $N=480$ ellipses (open symbols) and $240$ dimers (filled symbols) at
$\alpha = 1.0$ (circles), $1.002$ (squares), $1.01$ (diamonds), $1.05$
(upward triangles), $1.1$ (leftward triangles), $1.5$ (downward
triangles), and $2.0$ (rightward triangles).  The solid (dashed) line
has slope $1$ ($0.5$).  The dot-dashed lines have the form $z-z_J= 6.3
(\phi-\phi_J)/(\alpha-1)^{0.35}$.
\label{fig:dz_dphi}}
\vspace{-0.0in}
\end{figure}

\paragraph{Power-law scaling of linear shear modulus and contact number}
One of the hallmarks of the jamming transition in packings of
spherical particles is the power-law scaling of the static linear
shear modulus $G$ and contact number $z-z_J$ with $\phi-\phi_J$.  Both
scale as $(\phi-\phi_J)^{0.5}$ for linear repulsive
springs, which suggests that the contact number
scaling controls the behavior of the linear shear modulus~\cite{Ohern2003}.  In
Fig.~\ref{fig:G_dimers_ellipses}, we plot $G$ as a function of
$\phi-\phi_J$ for dimers (filled symbols) and ellipses (open symbols)
over a range of aspect ratios.  We again find power-law scaling near
jamming,
\begin{equation}
\label{static shear}
G = G_0(\alpha) (\phi-\phi_J)^{\beta},
\end{equation}
where $\beta=0.5$ and $G_0$ is weakly dependent on $\alpha$ for
dimers.  In contrast, $\beta=1$ for sufficiently small $\phi-\phi_J$
and $G_0(\alpha) \sim (\alpha-1)^{-0.44\pm 0.03}$
for ellipses.  The power-law scaling is stronger for ellipse packings,
and thus the ratio of the shear moduli $G_{\rm ellipse}/G_{\rm dimer}
\rightarrow 0$ in the limit $\phi \rightarrow \phi_J$ for all
$\alpha$. This implies that ellipse packings are much more susceptible
to shear in the linear response regime.

\begin{figure}
\epsfxsize=3.4in
\centerline{\epsffile{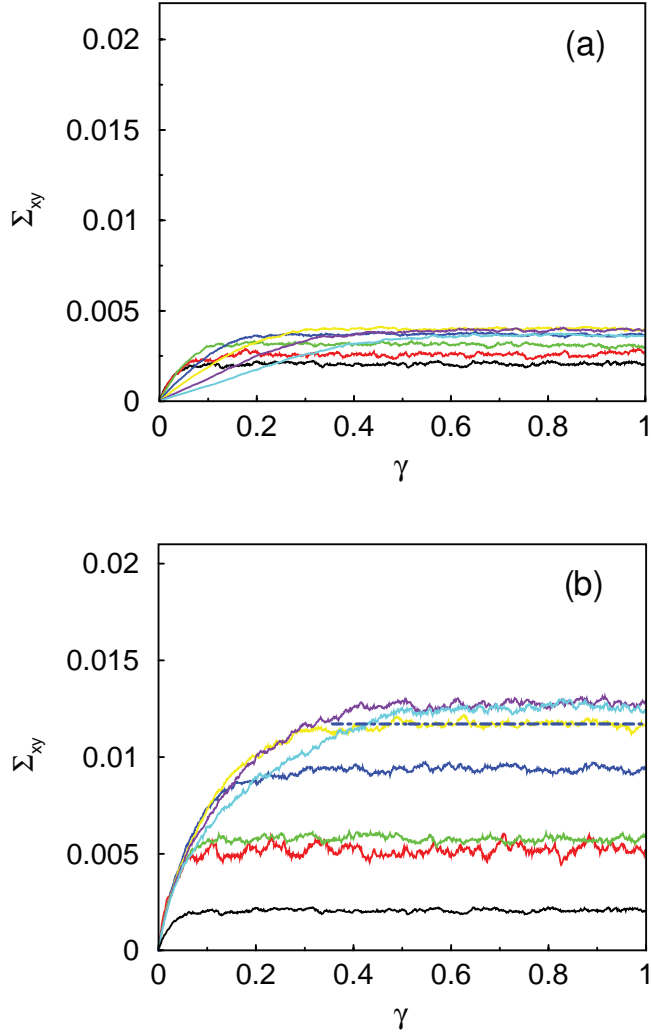}}
\vspace{-0.1in}
\caption{Shear stress $\Sigma_{xy}$ versus shear strain
$\gamma$ for packings of (a) ellipses and (b) dimers at $\phi-\phi_J =
10^{-1}$ for several aspect ratios $\alpha = 1.0$ (black), $1.05$
(red), $1.1$ (green), $1.2$ (blue), $1.3$ (yellow), $1.4$ (violet), and $1.5$
(cyan). The dashed horizontal line in (b) indicates the yield stress 
for dimer packings at $\alpha=1.3$.}
\label{fig:highphi}
\vspace{-0.0in}
\end{figure}

For jammed packings of spherical particles with linear spring
interactions both $G$ and $z-z_J$ scale as $(\phi-\phi_J)^{0.5}$.  We
find similar behavior, $G \sim z-z_J$, for dimer packings as a shown
in Fig.~\ref{fig:dz_dphi}.  For ellipse packings, we find
\begin{equation}
\label{contact}
z-z_J = z_0(\alpha) (\phi-\phi_J)^{\beta},
\end{equation}
where $z_0(\alpha) \sim (\alpha-1)^{-0.35\pm 0.1}$ and $\beta = 1$ for
sufficiently small $\phi - \phi_J$. Thus, $G$ and $z-z_J$ have the
same power-law scaling with $\phi-\phi_J$ even for hypostatic
packings.  We argued previously that the novel power-law scaling
exponent for $G$ and $z-z_J$ in ellipse packings originates from the
quartically stabilized vibrational modes~\cite{Mailman2009}.

\begin{figure}
\epsfxsize=3.2in
\centerline{\epsffile{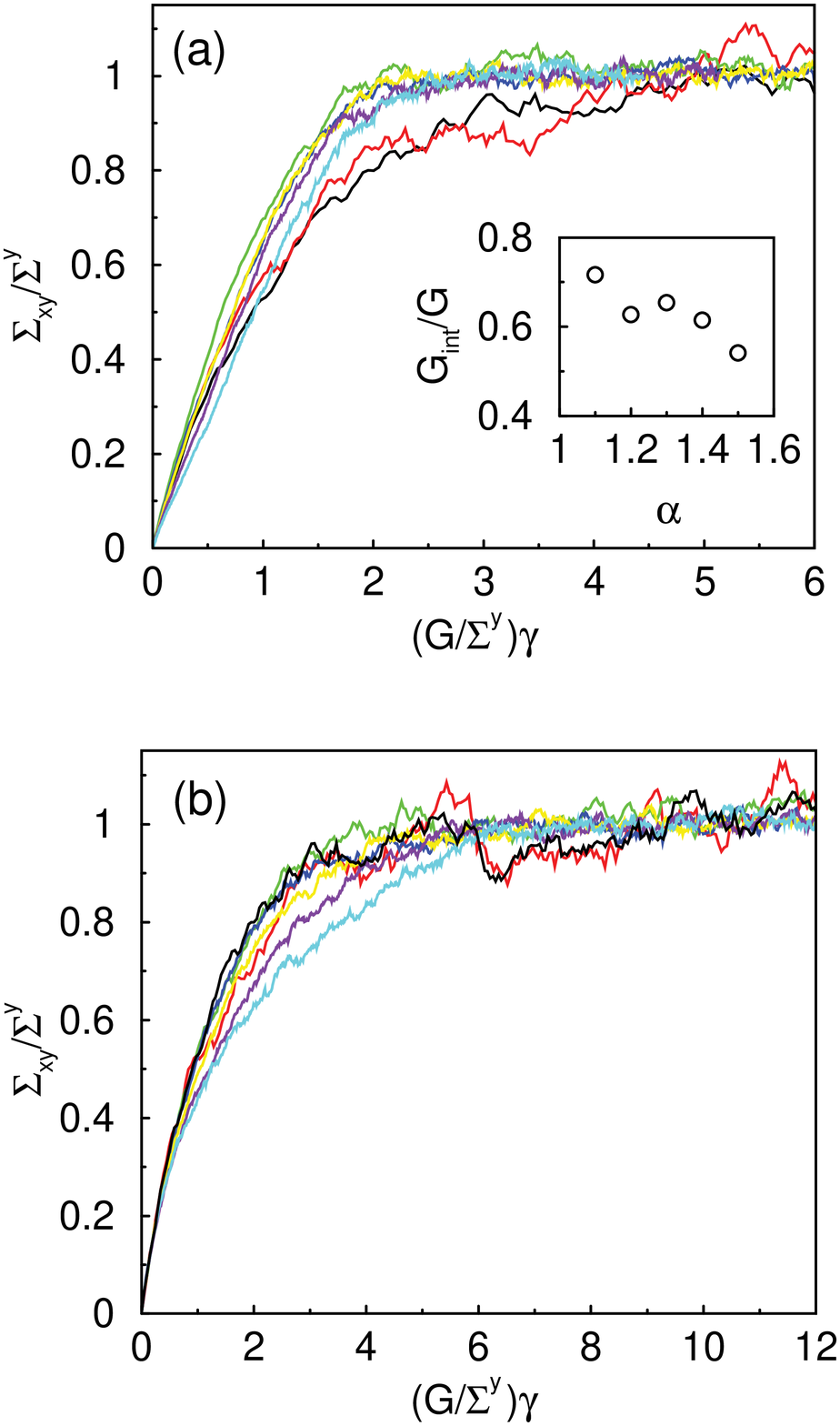}}
\vspace{-0.1in}
\caption{Same stress-strain relations in Fig.~\ref{fig:highphi} for (a)
ellipses and (b) dimers except the shear stress and strain have been
scaled by $\Sigma^y$ and $\Sigma^y/G$, respectively.  The inset to
panel (a) gives the average value of the slope ($G_{\rm int}/G$) of
the scaled stress-strain relation for shear stresses $\Sigma_{xy} <
\Sigma^y$ at aspect ratios $\alpha \geq 1.1$.}
\label{fig:highphi_scaled}
\vspace{-0.0in}
\end{figure}

\begin{figure}
\epsfxsize=3.7in
\centerline{\epsffile{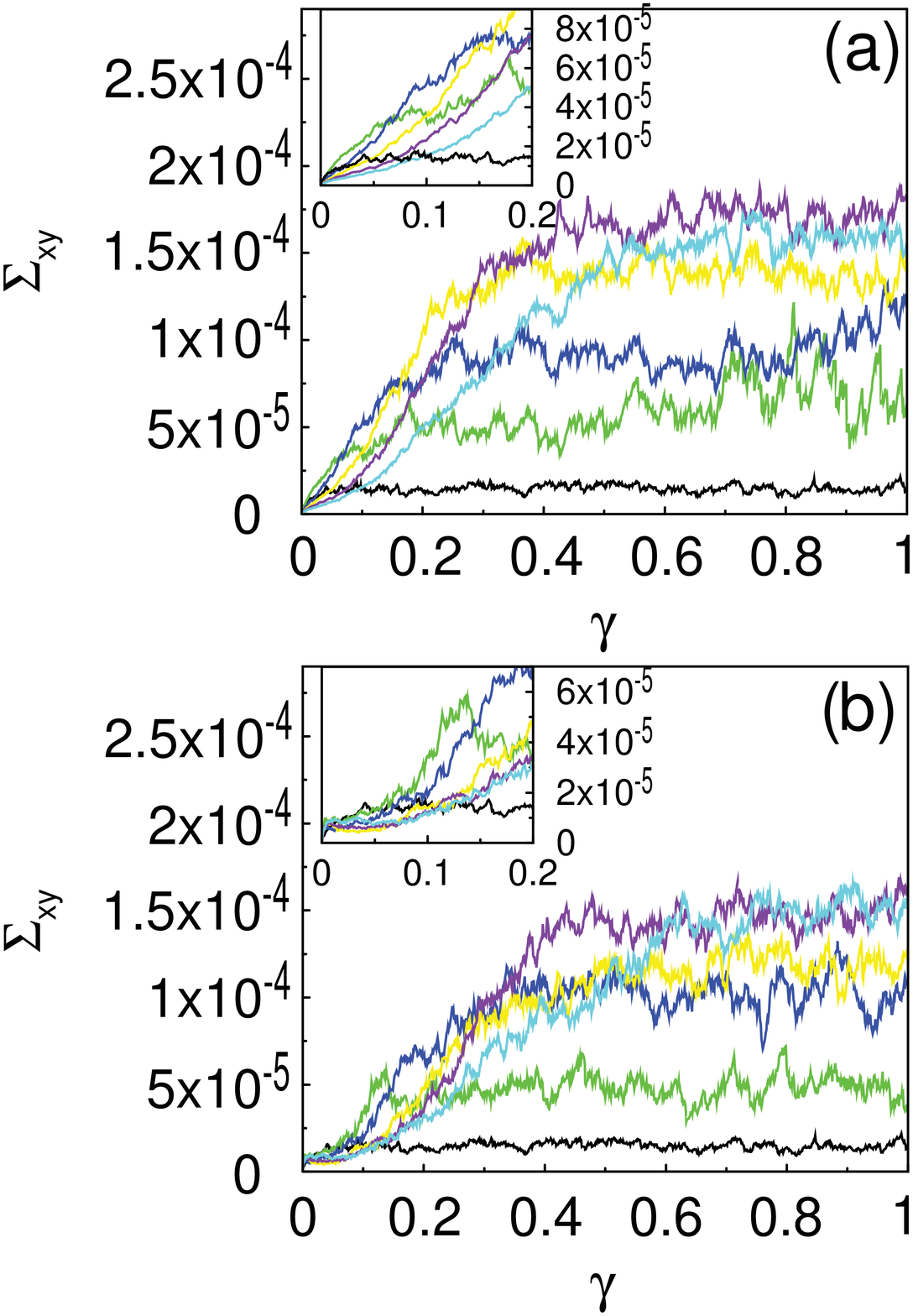}}
\vspace{-0.1in}
\caption{Shear stress $\Sigma_{xy}$ versus shear strain $\gamma$ for
packings of (a) ellipses and (b) dimers at $\phi-\phi_J = 10^{-3}$ for
several aspect ratios $\alpha = 1.0$ (black), $1.1$ (green), $1.2$
(blue), $1.3$ (yellow), $1.4$ (violet), and $1.5$ (cyan).  The insets
show the same data as in the main plots, except over a smaller range
of $\gamma$.
\label{fig:lowphi}}
\vspace{-0.0in}
\end{figure}

\begin{figure}
\epsfxsize=6.0in
\centerline{\epsffile{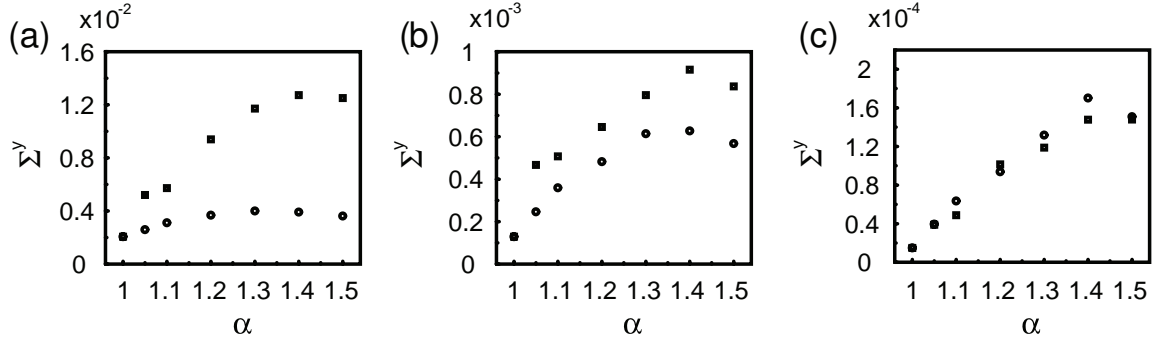}}
\vspace{-0.0in}
\caption{Yield stress $\Sigma^y$ as a function of aspect ratio
$\alpha$ at (a) $\phi - \phi_J=10^{-1}$, (b) $10^{-2}$, and (c)
$10^{-3}$ for ellipses (circles) and dimers (squares).}
\label{yield}
\vspace{-0.0in}
\end{figure}

\paragraph{Stress-strain relations} The full stress-strain behavior for 
ellipse and dimer packings is complex; it is qualitatively different
for ellipses and dimers and depends nontrivially on $\phi-\phi_J$ and
aspect ratio.  In Figs.~\ref{fig:highphi} and~\ref{fig:lowphi}, we
show the shear stress $\Sigma_{xy}$ versus strain $\gamma$ for
$\phi-\phi_J =10^{-1}$ and $10^{-3}$.  For ellipses at
$\phi-\phi_J=10^{-1}$ (Fig.~\ref{fig:highphi} (a)), the shear stress
is roughly linear with strain until the shear stress plateaus at the
yield stress, $\Sigma^y = \Sigma_{xy}(\gamma \rightarrow \infty)$,
which only weakly depends on aspect ratio and is at least a factor of
$2$ smaller than that for dimers ({\it c.f.} Fig.~\ref{yield} (a)).
We can achieve an approximate collapse of the stress-strain data for
ellipses at $\phi-\phi_J=10^{-1}$ for $\alpha \geq 1.1$ by scaling the
shear stress by $\Sigma_y$ and strain by $\Sigma_y/G$ as shown in
Fig.~\ref{fig:highphi_scaled} (a).  The inset to
Fig.~\ref{fig:highphi_scaled} (a) shows that the average shear modulus
defined over the wide range $0 \leq \Sigma_{xy} \leq \Sigma^y$ is
comparable to the linear response value, $G$, at small strains ({\it
c.f.} Fig.~\ref{fig:G_dimers_ellipses}) for $\alpha \geq 1.1$.

The behavior of the stress-strain curves for dimers at
$\phi-\phi_J=10^{-1}$ is qualitatively different from that for
ellipses as shown in Fig.~\ref{fig:highphi} (b). In particular, 
the approach of the shear stress to the yield stress plateau has 
significant curvature similar to the behavior found for 
sheared packings of spherical particles~\cite{ning}.  The scaled 
stress-strain curve in Fig.~\ref{fig:highphi_scaled} (b) emphasizes 
that dimer packings further strain soften as the aspect ratio increases.

Fig.~\ref{fig:lowphi} shows the stress-strain behavior for dimers and
ellipses much closer to the jamming transition at $\phi-\phi_J =
10^{-3}$.  At such small compressions, ellipse packings (panel (a)) no
longer possess such robust, sustained linear response over the range
$0 \leq \Sigma_{xy} \leq \Sigma^y$.  Instead, the shear stress is
first roughly linear with slope $\sim G$, but then stiffens on approach to
the yield stress.  However, the most striking feature of the
stress-strain curves at $\phi-\phi_J = 10^{-3}$ is the nearly perfect
{\it plastic} response (flat shear stress versus strain) for dimer
packings (panel (b)) with $\alpha \geq 1.1$.  The plastic regime
extends for strains from the end of the linear response regime to
$\gamma_p \approx 0.1$.  For $\gamma > \gamma_p$, the shear stress
grows rapidly as it approaches the yield stress, which is similar to
the behavior in Fig.~\ref{fig:highphi} (b) at $\phi-\phi_J=10^{-1}$.

\begin{figure}
\epsfxsize=5.5in
\centerline{\epsffile{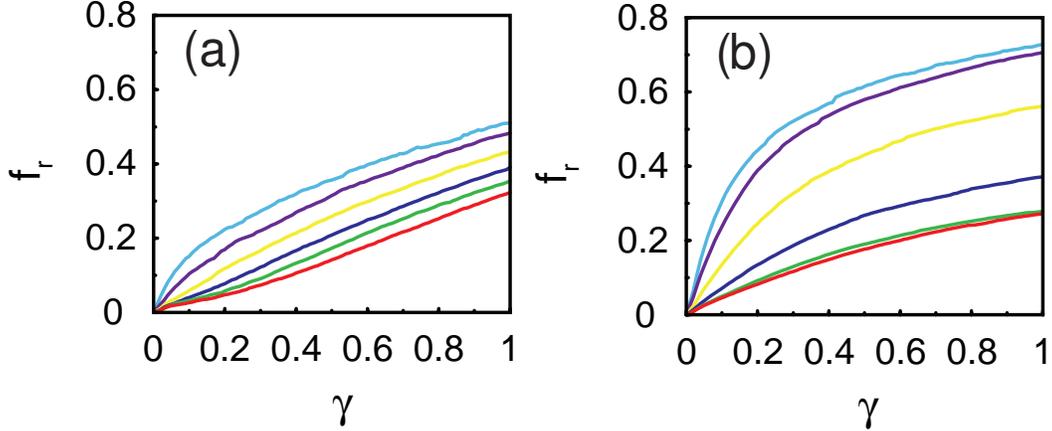}}
\vspace{-0.1in}
\caption{Fraction $f_r$ of the particle contacts at strain $\gamma$
that differ from those at $\gamma=0$ for (a) ellipses and (b) dimers
at $\alpha = 1.05$ (red), $1.1$ (green), $1.2$ (blue), $1.3$ (yellow),
$1.4$ (violet), and $1.5$ (cyan).}
\label{rearrangement}
\vspace{-0.0in}
\end{figure}

As demonstrated in the inset to Fig.~\ref{fig:lowphi} (a), ellipse
packings do not possess this nearly plastic response.  In the regime
$\gamma < \gamma_p$, the shear stress is roughly linear with a shear
modulus comparable to $G$. In Fig.~\ref{rearrangement}, we plot the
fraction of particle contacts at strain $\gamma$ that differ from
those at $\gamma=0$.  We find that the plastic behavior in dimer
packings is accompanied by a large increase in the number of particle
rearrangement events (changes in the contact network) over the strain
interval $0\leq \gamma \leq \gamma_p$.  Note that the largest fraction
(and rate of increase over $0\leq \gamma \leq \gamma_p$) of particle
rearrangements occurs for dimer packings with $\alpha = 1.5$, which
possess the most pronounced plastic response.  Further work is
required to elucidate the particle-scale motions that cause this
plastic response.

We have also calculated the nematic order parameter, $S = \langle \cos
[2 (\theta - \theta_0)] \rangle$, where $\theta_0$ is average
orientation of the particles, as a function of shear strain.  We find
that the nematic order increases roughly linearly with $\gamma$ up to
strains of $0.2-0.3$ in both sheared dimer and ellipse packings.  Thus, it
is possible that nematic order leads to qualitatively different effects in
dimer and ellipse packings since dimers display plastic response,
while ellipses do not.

In Fig.~\ref{yield}, we show the yield stress $\Sigma^y$ for dimers
and ellipses as a function of aspect ratio at (a)
$\phi-\phi_J=10^{-1}$, (b) $10^{-2}$, and (c) $10^{-3}$.  We find that
the yield stress increases with aspect ratio, which acts as an
effective friction coefficient~\cite{leo}. However, $\Sigma^y$ begins
to level off near $\alpha^* \sim 1.4$, which is likely related to a
maximum in the nematic order near $\alpha^*$. In contrast to the
behavior at small shear strains, the yield stress for dimers and
ellipses becomes nearly identical near jamming at
$\phi-\phi_J=10^{-3}$.  Thus, measurements of the jamming packing fraction
$\phi_J$ and yield stress $\Sigma^y$ are relatively insensitive to 
microscale geometrical features of individual particles.

\section{Future Directions}
\label{future}

These studies of the structural and mechanical properties of dimer and
ellipse packings raise a number of interesting questions that will
likely spur new research activity in this area.  First, there are many
features of jammed ellipse and dimer packings that are different.
Most notably, ellipse packings are hypostatic while dimer packings are
isostatic, which gives rise to novel power-law scaling of the
structural and mechanical properties near jamming.  Thus, we have
shown that the macroscopic jamming behavior of anisotropic particles
depends sensitively on the microscale geometrical features of
individual particles.  

However, the dependence on aspect ratio of the jamming packing
fraction and yield stress, which strongly affect the glass transition
in thermalized systems~\cite{michele}, is relatively insensitive to
whether the packings are composed of dimers or ellipses. Thus, it is
not clear {\it a priori} which structural, mechanical, and dynamical
properties are sensitive to microscale particle properties.  Thus, we
encourage reinvigorated studies of atomic, colloidal, and granular
systems to determine under what circumstances geometrical features of
individual particles play an important role in jamming behavior and
glassy dynamics.

\section*{Acknowledgments}

This work was supported by the National Science Foundation under Grant
Nos. CAREER-DMR0448838 and DMS0835742.  The authors are grateful to B.
Chakraborty, A. Donev, E. Dufresne, M. Mailman, K. Schweizer, and
S. Torquato for helpful discussions.  We also thank the Yale High
Performance Computing Center for providing the required computing
resources.

\appendix

\section{Packing fraction}
\label{packing_fraction}

In our numerical simulations, we consider bidisperse mixtures of
dimers and ellipses in which one-third ($N_l=1/3$) of the particles
are large (with a minor axis $1.4$ times that of the smaller
particles, {\it i.e.} $b_l = 1.4 b_s$) and two-thirds ($N_s=2/3$) of
the particles are small.  When calculating the packing fraction for
rigid dimers (fused disks), we do not double count the overlapping
region.  Thus, we define the packing fraction for dimers in $2D$ as
\begin{equation}
\phi_{\rm dimer} = 2 N_s \pi \left(\frac{b_s}{L}\right)^2 \left(1
+ \frac{N_l}{N_s} \left( \frac{b_l}{b_s}\right)^2 \right) \left(1 -
\frac{1}{\pi}\left[\cos^{-1}(\alpha-1) + (\alpha-1) \sqrt{\alpha
(2-\alpha)}\right]\right).
\end{equation}
For ellipses
\begin{equation}
\phi_{\rm ellipse} = N_s \pi \alpha \left(\frac{b_s}{L}\right)^2 \left(1 + \frac{N_l}{N_s}
\left( \frac{b_l}{b_s}\right)^2 \right).  
\end{equation}

\section{Contact distance}
\label{contactdistance}

The Perram and Wertheim formulation for calculating the contact
distance $\sigma_{ij}$ between ellipses $i$ and $j$ with
orientations ${\hat \mu}_i$ and ${\hat \mu}_j$ and center-to-center
direction ${\hat r}_{ij}$ involves the following minimization
procedure~\cite{Perram}:
\begin{equation}
\label{sigma_functional}
\sigma_{ij} = \min_\lambda \sigma(\lambda) = \min_\lambda
\frac{\sigma_0(\lambda)}
{\sqrt{1-\frac{\chi(\lambda)}{2}\displaystyle\sum_{\pm}
\frac{(\beta(\lambda)\hat{r}_{ij}\cdot\hat{\mu}_i\pm
\beta(\lambda)^{-1}\hat{r}_{ij}\cdot\hat{\mu}_j)^2}
{1\pm\chi(\lambda)\hat{\mu}_i\cdot\hat{\mu}_j}}},
\end{equation} 
where 
\begin{equation}
\sigma_0(\lambda)
=\frac{1}{2}\left(\frac{b_i^{2}}{\lambda}+\frac{b_j^{2}}
{1-\lambda}\right)^{1/2},
\end{equation}
\begin{equation}
\chi(\lambda)
=\left(\frac{\left(a_{i}^{2}-b_{i}^{2}\right)\left(a_{j}^{2}-
b_{j}^{2}\right)} {\left(a_{j}^{2}+\frac{1-\lambda}{\lambda}
b_{i}^{2}\right)
\left(a_{i}^{2}+\frac{\lambda}{1-\lambda}b_{j}^{2}\right)}\right)^{1/2},
\end{equation}
and
\begin{equation}
\beta(\lambda)
=\left(\frac{\left(a_i^2-b_i^2\right)\left(a_j^2+\frac{1-\lambda}{\lambda}b_i^2\right)}
{\left(a_j^2-b_j^2\right)\left(a_i^2+\frac{\lambda}{1-\lambda}b_j^2\right)}\right)^{1/4}.
\end{equation}
Determining $\lambda_{\rm min}$ that minimizes $\sigma(\lambda)$
(Eq.~\ref{sigma_functional}) involves solving for the roots of a
quartic polynomial in $\lambda$ for $2D$ bidisperse
systems~\cite{Schreck2009}.  The polynomials can be expressed
analytically in terms of ${\hat \mu}_i$, ${\hat \mu}_j$, ${\hat
r}_{ij}$, and the major and minor axes of particles $i$ and $j$, and
then solved using Newton's method.

\section{Packing-generation algorithm}
\label{packing}

In Sec.~\ref{methods}, we outlined our procedure to generate
static packings of dimers and ellipses.  Here, we provide
some of the numerical parameters involved in the simulations.  For the
energy minimization, we employ the conjugate gradient
technique~\cite{numrec}, where the particles are treated as massless.
The two stopping criteria for the energy minimization are $V_{t} - V_{t-1} <
V_{\rm tol} = 10^{-12}$ and $V_t < V_{\rm min} = 10^{-12}$, where
$V_t$ is the potential energy per particle at iteration $t$, and the
target potential energy per particle of a static packing is $V_{\rm tol}
< V/N < 2 V_{\rm tol}$.  For the first compression or decompression
step we use the packing-fraction increment $\Delta\phi = 10^{-3}$.
Each time the procedure switches from expansion to contraction or vice
versa, $\Delta \phi$ is reduced by a factor of $2$.  Using the packing
generation procedure with these parameters, we are able to locate the
jamming threshold in packing fraction $\phi_J$ to within $10^{-6}$ for
each static packing.  

\section{Calculation of forces and torques}
\label{torque}

In this appendix, we provide specific details for calculating the
interparticle forces and torques for dimers and ellipses, which are
required to perform energy minimization and evaluate the shear
stress. The forces and torques can be obtained from the interaction
potential (Eq.~\ref{interaction}) using generalizations of ${\vec
F}_{ij} = dV/d{\vec r}_{ij}$, where ${\vec F}_{ij}$ is the force on
particle $i$ due to particle $j$.

\paragraph{Dimers} For dimers, the
interaction force on monomer $k_i$ belonging to dimer $i$ from monomer $k_j$
belonging to a distinct dimer $j$ is
\begin{equation}
{\vec F}_{k_i,k_j} = \frac{dV}{d{\vec r}_{k_i,k_j}}.
\end{equation}
The total force on dimer $i$ is obtained by summing over all monomers
$k_i$ belonging to dimer $i$, all dimers $j$ different from $i$, and all
monomers $k_j$ belonging to dimer $j$:
\begin{equation}
{\vec F}_i = \sum_{k_i} \sum_j \sum_{k_j} {\vec F}_{k_i,k_j}. 
\end{equation}

The torque on dimer $i$ arising from an interaction between monomer 
$k_i$ on dimer $i$ and monomer $k_j$ belonging to dimer $j$ is given 
by 
\begin{equation}
\label{dimer_torque}
{\vec T}_{k_i,k_j} = {\vec r}_{k_i} \times {\vec F}_{k_i,k_j},
\end{equation}
where ${\vec r}_{k_i}= d_i (\cos \theta_i {\hat x} - \sin \theta_i
{\hat y})$ is the vector from the center of dimer $i$ to the center of
monomer $k_i$, $d_i = b_i (\alpha - 1)/2$, and $\theta_i$ gives the
orientation of dimer $i$. The total torque on dimer $i$, ${\vec
T}_i$, is obtained by summing ${\vec T}_{k_i,k_j}$ over all monomers
$k_i$ on dimer $i$, all dimers $j$ distinct from $i$, and all monomers
$k_j$ on dimer $j$.

\begin{figure}
\epsfxsize=5.0in
\centerline{\epsffile{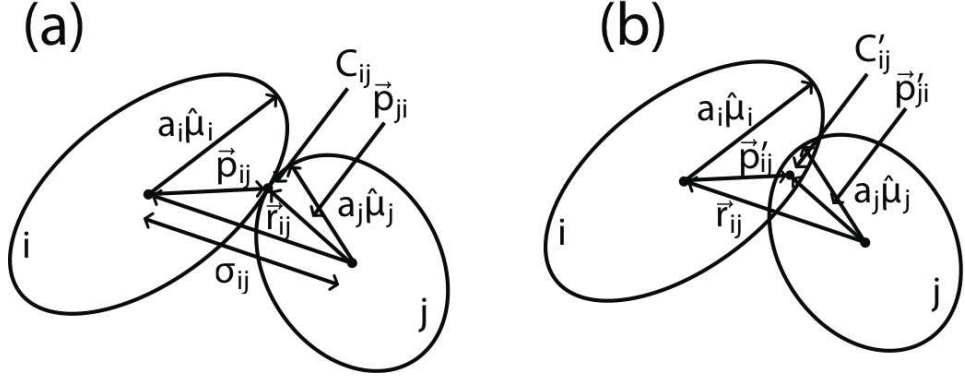}}
\vspace{0.0in}
\caption{\label{fig:contactpoint} Definition of the point of contact
$C_{ij}$ for ellipses $i$ and $j$ that are (a) `just touching' and (b)
overlapped.  ${\vec p}_{ij}$ is location of the point of contact 
relative to the center of mass of ellipse $i$.  In the overlapped 
case, the effective point of contact $C^{'}_{ij}$ is given by ${\vec p}_{ij}'
= {\vec p}_{ij} (r_{ij}/\sigma_{ij})$. }
\vspace{-0.00in}
\end{figure}

\paragraph{Ellipses}
For ellipses $i$ and $j$, the interparticle force depends explicitly
on how the contact distance $\sigma_{ij}$ varies with the vector
separation ${\vec r}_{ij}$:
\begin{equation}
\label{force_ellipse}
F_{\xi ij} 
= -\frac{\partial V}{\partial r_{ij}}\left(\frac{\xi_{ij}}{r_{ij}}-
\frac{r_{ij}}{\sigma_{ij}}
\frac{\partial\sigma_{ij}}{\partial\xi_{ij}}\right),
\end{equation} 
where $\xi = x,y$.  To calculate the torque, one must specify the
point of contact. For `just touching' ellipses $i$ and $j$ are in contact
at only one point, as shown in Fig.~\ref{fig:contactpoint} (a), the
location ${\vec p}_{ij}$ of the point of contact $C_{ij}$ (relative to
the center of mass of ellipse $i$) is unambiguous and given by
\begin{eqnarray}
\label{poc}
\vec{p}_{ij}&=&p_{ij}^0(\cos(\psi_{ij}+\theta_i)\hat{x}+\sin(\psi_{ij}+\theta_i)\hat{y})\\
p_{ij}^0&=&\frac{1}{2\sqrt{ \left( \frac{\cos \psi_{ij}}{a_i}\right)^2 + \left( \frac{\sin \psi_{ij}}{b_i} \right)^2 }}\\
\tan \psi_{ij} &=&\frac{1}{\alpha^2}
\frac{\tan(\Theta_{ij}-\theta_i)-\sigma_{ij}^{-1} \frac{\partial\sigma_{ij}}{\partial\beta_{ij}}}
{1+\sigma_{ij}^{-1} \tan(\Theta_{ij}-\theta_i)\frac{\partial\sigma_{ij}}{\partial\Theta_{ij}}},
\end{eqnarray}
where $\cos \beta_{ij} = {\hat \mu}_i \cdot {\hat r}_{ij}$, $\cos \Theta_{ij}={\hat x}_i \cdot {\hat r}_{ij}$. The torque $T_{ij}$ on
ellipse $i$ from $j$ is then
\begin{equation}
\label{ellipse_torque}
T_{ij}=p_{x ij}F_{y ij}-p_{y ij}F_{x ij}.
\end{equation} 
As shown in Fig.~\ref{fig:contactpoint} (b), upon compression,
ellipses are no longer `just touching', and thus Eq.~(\ref{poc}) for
the point of contact $C_{ij}$ is no longer exact. In this case, we
scale ${\vec p}_{ij}$ by $r_{ij}/\sigma_{ij}$, which yields an
effective point of contact $C^{'}_{ij}$ that is within the overlap
region of the two ellipses.

\section{Calculation of shear stress}
\label{shear}

For systems composed of spherical particles, the correct form 
for the stress tensor ${\hat \Sigma}_{\alpha \beta}$ in $2D$, where 
$\alpha, \beta=x,y$, is the virial expression~\cite{allen}: 
\begin{equation}
\label{virial}
{\hat \Sigma}^V_{\alpha \beta} = \frac{1}{2 L^2} \sum^{N}_{i>j=1} \left(
F_{ij\alpha} r_{ij\beta} + F_{ij\beta} r_{ij\alpha} \right),
\end{equation}
where $F_{ij\alpha}$ is the $\alpha$-component of the force ${\vec
F}_{ij}$ on particle $i$ arising from an overlap with particle $j$,
$r_{ij\beta}$ is the $\beta$-component of the vector ${\vec r}_{ij}$
from the center of mass of particle $j$ to that of
particle $i$.

The correct form for the stress tensor ${\hat \Sigma}_{\alpha \beta}$
in $2D$ for systems composed of {\it anisotropic} particles is the
Love expression~\cite{edwards}:
\begin{equation}
\label{love}
{\hat \Sigma}^L_{\alpha \beta} = \frac{1}{2 L^2} \sum^{N}_{i,j=1} \left(
F_{ij\alpha} p_{ij\beta} + F_{ij\beta} p_{ij\alpha} \right),
\end{equation}
where $p_{ij\beta}$ is the $\beta$-component of the vector from the center
of mass of particle $i$ to the point of contact $C_{ij}$ with particle
$j$.  Note that the Love expression reduces to the virial expression 
for spherical particles.

\begin{figure}
\epsfxsize=4.0in
\centerline{\epsffile{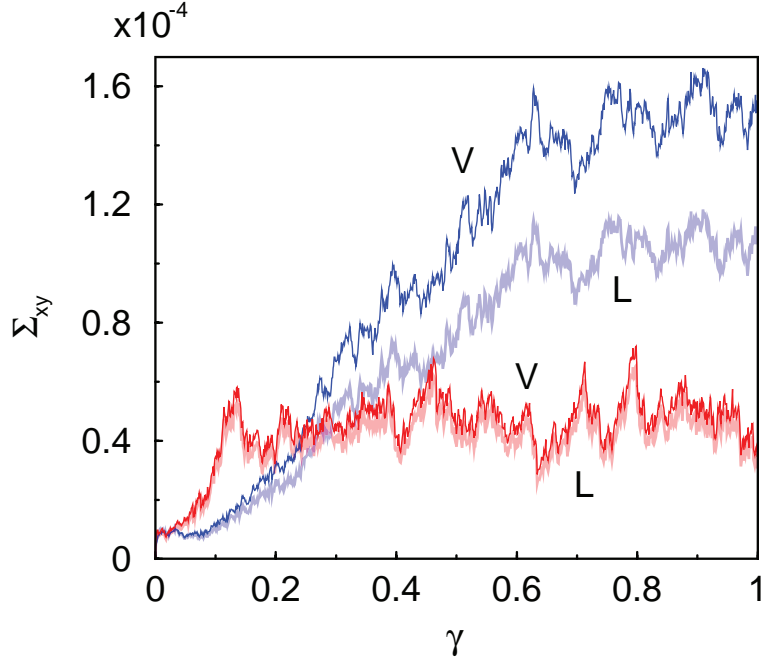}}
\vspace{0.0in}
\caption{\label{love_virial} Comparison of the Love (L) and
virial (V) expressions for the shear stress $\Sigma_{xy}$
as a function of shear strain $\gamma$ for aspect ratio $\alpha=1.1$
(red) and $1.5$ (blue) at $\phi-\phi_J=10^{-3}$.}
\vspace{-0.00in}
\end{figure}

In our studies of simple shear, we focus on the off-diagonal component
of the stress tensor $\Sigma_{xy}$. Calculating the point of contact
at each shear strain is computationally expensive; we have therefore
used the virial expression $\Sigma^V_{xy}$ instead of the Love
expression $\Sigma^L_{xy}$ to quantify the shear stress for both dimer
and ellipse packings.  As a check, we measured both $\Sigma^L_{xy}$ and
$\Sigma^V_{xy}$ for dimers as a function of aspect ratio and
compression.  Fig.~\ref{love_virial} shows that they give
quantitatively similar results for $\alpha=1.1$ and $\alpha=1.5$ (for
$\gamma < 0.2$), and qualitatively similar results for $\alpha=1.5$ at
large strain.  In particular, the plastic response of dimer packings 
at small compressions is unaffected by the choice of the definition 
of the shear stress.

\end{document}